\documentclass[bibyear]{aa} 
\usepackage{graphicx,rotate}
\usepackage{txfonts}
\usepackage{booktabs}

\newcommand{\vm}{V_{\rm max}}

\newcommand{\rd}{r_{\rm d}}
\newcommand{\oii}{[O\,{\small II}]}
\newcommand{\nii}{[N\,{\small II}]}
\newcommand{\oiii}{[O\,{\small III}]}
\newcommand{\ha}{H$\alpha$}
\newcommand{\sg}{\sigma_{\rm g}}
\newcommand{\clusters}{(pro\-to-)clus\-ters}
\newcommand{\cluster}{(pro\-to-)clus\-ter}

\begin{document} 

\title{Kinematics of disk galaxies in \clusters\ at $z=1.5$
\thanks{Based on observations with the European Southern Observatory
Very Large Telescope (ESO-VLT), observing run ID 099.B-0644A.}
}

\author{
A.~B\"ohm\inst{1},
B.~L.~Ziegler\inst{1},
J.~M.~P\'erez-Mart\'inez\inst{1,2},
T.~Kodama\inst{2},
M.~Hayashi\inst{3},
C.~Maier\inst{1},
M.~Verdugo\inst{1},
Y.~Koyama\inst{4,5}
}

\institute{Institute for Astronomy (IfA), University of Vienna, T\"urkenschanzstrasse 17, 1180 Vienna, Austria
\email{asmus.boehm@univie.ac.at}
\and
Astronomical Institute, Tohoku University, Aramaki, Aoba-ku, Sendai 980-8578, Japan
\and
National Astronomical Observatory of Japan, Osawa, Mitaka, Tokyo 181-8588, Japan
\and
Subaru Telescope, National Astronomical Observatory of Japan, 650 North A’ohoku Place, Hilo, HI 96720, USA
\and
Department of Astronomical Science, SOKENDAI (The Graduate University for Advanced Studies), Mitaka, Tokyo 181-8588, Japan
}

\date{Received 24 March 2019 / Accepted 4 December 2019}

 
\abstract
{} 
{
While many aspects of the impact of dense environments on late-type galaxies at redshifts below unity have been scrutinized in the past decades, observational studies of the interplay between environment and disk galaxy evolution at $z>1$ are still scarce. We observed star-forming galaxies at $z \approx 1.5$ selected from the HyperSuprimeCam Subaru Strategic Program. The galaxies are part of two significant overdensities of \oii\ emitters identified via narrow-band imaging and photometric redshifts from $grizy$ photometry.
}
{
We used VLT/KMOS to carry out \ha\ integral field spectroscopy of 46 galaxies in total. Ionized gas maps, star formation rates and velocity fields were derived from the \ha\ emission line. We quantified morphological and kinematical asymmetries to test for potential gravitational (e.g.~galaxy-galaxy) or hydrodynamical (e.g.~ram-pressure) interactions.
}
{
\ha\ emission was detected in 36 of our targets. 34 of the galaxies are members of two \clusters\ at $z=1.47$, confirming our selection strategy to be highly efficient. 
By fitting model velocity fields to the observed ones, we determined the intrinsic maximum rotation velocity $\vm$ of 14 galaxies. Utilizing the luminosity-velocity (Tully-Fisher) relation, we find that these galaxies are more luminous than their local counterparts of similar mass by up to $\sim$\,4\,mag in the rest-frame $B$-band. In contrast to field galaxies at $z<1$, the offsets of the $z \approx 1.5$ \cluster\ galaxies from the local Tully-Fisher relation are not correlated with their star formation rates but with the ratio between $\vm$ and gas velocity dispersion $\sigma_{\rm g}$. This probably reflects that, as is observed in the field at similar redshifts, fewer disks have settled to purely rotational kinematics and high $\vm/\sigma_{\rm g}$ ratios. 
Tests with degraded low-redshift cluster galaxy data show that we can not identify purely hydrodynamical interactions with the imaging currently at hand. Due to relatively low galaxy velocity dispersions ($\sigma_{\rm v}<400$\,km/s) of the \clusters, gravitational interactions likely are more efficient, resulting in higher kinematical asymmetries, than in present-days clusters.
}
{}


\titlerunning{Disk galaxies in $z=1.5$ \clusters}
\authorrunning{A. B\"ohm et al.}

\maketitle

\section{\label{intro}Introduction}

The evolution of galaxies is affected by their environment in various ways. Observations show that the galaxy population in clusters has changed substantially during the past 5-6 Gyr, with the fraction of spirals decreasing and the fraction of lenticulars (but also dwarf ellipticals) increasing towards $z \approx 0$ (e.g.~Desai et al.~\cite{desa07}). Galaxies in dense regions on average have redder colors (e.g.~Blanton et al.~\cite{blan05}) and are less frequently star-forming (e.g.~Verdugo et al.~\cite{verd08}) than in the field. In the cluster environment, galaxies are subject to a plethora of interaction processes, such as harassment (e.g.~Moore et al.~\cite{moor96}), ram-pressure stripping (RPS, e.g.~Kronberger et al.~\cite{kron08}), or strangulation (e.g.~Balogh et al.~\cite{balo00}). The latter two refer to the impact of the hot intra-cluster medium (ICM) on the interstellar medium of a galaxy moving within a cluster. Strangulation occurs when only the gaseous halo of a galaxy is removed, followed by a phase of gas consumption via star formation on a time scale of Gyrs. 
RPS, first discussed by Gunn \& Gott~(\cite{gunn72}), affects not only the hot gas halo but also the cold gas disk. Quilis, Moore and Bower~(\cite{quil00}) have shown with hydrodynamical simulations that ram-pressure plays an important role in the transformation of field spirals into cluster S0 galaxies. More recently, e.g.~Steinhauser et al.~\cite{stei16}) demonstrated that, in extreme cases, RPS can quench star formation within $\sim$\,0.5\,Gyr.

As part of the STAGES survey (Gray et al.~\cite{gray09}) of the multiple cluster system A901/902 at $z=0.17$, we were able to show in B\"osch et al.~(\cite{boes13a}) that so-called red spirals are probably produced by the impact of ram-pressure stripping. Red spirals, which show weaker spiral arms and a four times lower specific star formation rate than normal, blue spirals, might be an intermediate stage in the transformation of blue field spirals into cluster S0s. Red spirals are the dominant population in clusters at intermediate cluster-centric radii and galaxy masses, while they are almost absent in the field at $z\,\approx\,0.2$ (e.g.~Wolf et al.~\cite{wolf09}). In B\"osch et al.~(\cite{boes13b}), we used the Tully-Fisher relation (TFR; Tully \& Fisher~\cite{tull77})~--- the scaling relation linking the maximum rotation velocity $\vm$ of disk galaxies to their luminosity~--- to confirm that red spirals are in the process of quenching and that the TFR scatter is increasing towards the inner cluster regions. Other studies investigated the mass-size relation, e.g.~finding smaller sizes of star-forming galaxies in clusters at $z\,\approx\,0.5$ than in the field (Kuchner et al.~\cite{kuch17}).

In the past decades, a large number of studies have utilized the TFR. Using field spirals at $0.1<z<1.0$, B\"ohm \& Ziegler~(\cite{boeh16}) found that, at given $\vm$, disk galaxies $\sim$\,8\,Gyr ago were brighter by 1.2\,mag in rest-frame $B$ and smaller by a factor $\sim$\,1.5 than spirals in the present-day universe. Based on the stellar-mass TFR, Miller et al.~(\cite{mill12}) found only a very small evolution of its zero point corresponding to 0.06\,dex smaller stellar masses at $z\approx 1.7$ and fixed $\vm$. While these two former studies relied on slit spectroscopy, the usage of Integral Field Units like VLT/KMOS has become more common for kinematic studies of distant galaxies. \"Ubler et al.~(\cite{uebl17}), e.g., found no evolution in the stellar-mass TFR up to $z \approx 2.3$ using KMOS.

For several years, the results on whether the TFR differs between the field and the cluster environment were somewhat heterogeneous. Ziegler et al.~(\cite{zieg03}) or Nakamura et al.~(\cite{naka06}), e.g., did not find differences between the two regimes, while Milvang-Jensen et al.~(\cite{milv03}) or Bamford et al.~(\cite{bamf05}) found that late-type galaxies in dense environments are more luminous than in the field. Other studies, like Moran et al.~(\cite{mora07}) deduced that the TFR scatter in clusters is higher than in low-density environments. It became clearer later that one key factor in such kinematical studies is to compare only galaxies with similar properties. Since various cluster-specific interactions can affect the kinematics of cluster disk galaxies, the fraction of disturbed rotation curves or velocity fields is found to be higher in clusters than in the field (e.g. Vogt et al. \cite{vogt04}). This can introduce an environmental dependence of the TFR unless the same criteria are applied to all kinematic data by using only symmetric rotation velocity fields in the TFR analysis (e.g. B\"osch et al.~\cite{boes13b}). 

Galaxy clusters at high redshifts $z>1$ show important differences to those at lower redshifts. While galaxies in the central regions of low-redshift clusters mostly are passive, clusters at high-$z$ frequently show strongly star-forming galaxies in their central regions. E.g., Hayashi et al.~(\cite{haya10}) found a high number density of \oii\ emitters in the core of XMMXCS\,J2215.9-1738 at $z=1.46$. Some clusters already at these early cosmic epochs show a well-established intra-cluster medium (ICM) in hydrostatic equilibrium (e.g.~IDCS1426+358 at $z=1.75$, Brodwin et al.~\cite{brod16}). TFR studies in $z>1$ clusters still are scarce. Based on a small sample of disk galaxies in a $z=1.4$ cluster, we found a moderate luminosity evolution of high-mass cluster galaxies, while low-mass cluster galaxies where much brighter than their field counterparts at similar redshifts (P\'erez-Mart\'inez et al.~\cite{pere17}).

Multiple observations have shown that, even in the absence of interaction processes typical for the cluster regime, disk galaxies become more kinematically hot towards higher redshifts in the sense that the ratio between their gas maximum rotation velocity $\vm$ and velocity dispersion $\sigma_{\rm v}$ decreases. The settling of disks, i.e. $\vm/\sigma_{\rm v}$ ratios above a certain threshold, occurs at earlier cosmic epochs for disks of higher mass (e.g.~Kassin et al.~\cite{kass12}). As Simons et al.~(\cite{simo16}) have found at redshifts $z \approx 2$, this behavior is also reflected in the stellar-mass TFR, where galaxies with low $\vm/\sigma_{\rm v}$ ratios tend to show too slow rotation for their stellar mass.

In this paper, we aim to shed light onto the impact of dense environments on the kinematic evolution of disk galaxies nine Gyr ago. The paper is organized as follows: In Sect.~\ref{obs} we outline the target selection and observations, Sect.~\ref{data} briefly describes the data reduction, in Sect.~\ref{ana} we detail all steps of the analysis, Sect.~\ref{disc} comprises a discussion and Sect.~\ref{concl} summarizes our main results.

In the following, we assume a flat concordance cosmology with $\Omega_\Lambda=0.7$, $\Omega_m=0.3$ and $H_0=70$\,km\,s$^{-1}$\,Mpc$^{-1}$. All magnitudes are given in the Vega system.

\section{\label{obs}Target Selection and Observations}

Our target selection relied on the HyperSuprime-Cam Subaru Strategic Program (HSC-SSP, see Aihara et al.~\cite{aiha18a}). The HSC (Miyazaki et al.~\cite{miya18}, Komiyama et al.~\cite{komi18}, Kawanomoto et al.~\cite{kawa18}, Furusawa et al.~\cite{furu18}) is an optical imaging camera operated at the prime focus of Subaru that, with a diameter of 1.5$^\circ$, features the largest field-of-view of all 10\,m-class telescopes. HSC offers five broad-band filters, $g$, $r$, $i$, $z$, $y$, and several narrow-band filters of which the filter NB921 probes the 3727\,\AA\ \oii\ emission line doublet of galaxies at a redshift $z \approx 1.5$ (Hayashi et al.~\cite{haya18}). The HSC-SSP spans a total of 300 nights at Subaru over 5-6 years. The first data release took place 1.7\,years into the survey; it is described in Aihara et al.~(\cite{aiha18b}).

The two target \clusters\ for our spectroscopic follow-up, which we will refer to as HSC-CL2329 and HSC-CL2330 in the following, were identified as strong overdensities (with a significance of 5.7\,$\sigma$ and 7.2\,$\sigma$, resp.) of \oii\ emitters at $z=1.47$ using the narrow-band filter NB921. Clusters or \clusters\ at these redshifts often show star-forming galaxies in their central regions, while they are preferentially located between intermediate cluster-centric radii and the outskirts of \emph{local} clusters (Dressler~\cite{dress80}). Since HSC-SSP also comprises photometric redshifts (derived via various methods, including SED fitting, machine learning, etc., see Tanaka et al.~\cite{tana18}), we could rule out any contamination by other emission lines, e.g. \ha\ or \oiii, from  galaxies at different redshifts. This was later confirmed by the spectroscopic redshift distribution of our sample (see Sect.~\ref{ana}). Since the width of the NB921 filter corresponds to a redshift range of $z=1.471\pm0.018$, or a velocity width of $\pm$2200\,km/s, we were expecting that i) the majority of the \oii\ emitters are physically associated \cluster\ members and ii) the \cluster\ sample will be nearly complete within the covered field-of-view, given that the typical velocity dispersion of rich clusters is 1000\,km/s.

The K-band Multi Object Spectrograph (KMOS) is a second-generation VLT instrument capable of observing 24 science targets simultaneously in the near-infrared. Each of the individual IFUs offers a field-of-view of 2.8\,$\times$\,2.8\,arcsec$^2$ (with spaxels sizes of 0.2\,arcsec),
corresponding to 23.7\,kpc on a side at $z=1.47$, 
and can be placed within a patrol field of 7.2\,arcmin diameter. In the $H$-band, which we chose for our observations, the typical spectral resolution of KMOS is $R \approx 4000$.

We selected our target galaxies to have \ha\ fluxes 
$f_{\rm H\alpha} > 6 \times 10^{-17}$\,erg\,s$^{-1}$\,cm$^{-2}$. 
Predicted \ha\ fluxes were converted from the observed \oii\ fluxes following Kennicutt~(\cite{kenn98}) assuming the \ha\ fluxes to be on average a factor of two larger than the \oii\ fluxes. 
The KMOS $H$-band observations were carried out between June and September 2017 with a stare / nod-to-sky strategy. Each Observation Block consisted of an ABA ABA sequence, where ``A'' denotes that the IFUs were placed on the science targets, and ``B'' indicates that the IFUs were observing blank sky. The integration time of each exposure was 420\,s, the total integration time per \cluster\ was $\sim$\,2.8\,h and the total observing time per \cluster, including overheads, was four hours. Seeing conditions measured in the optical with the DIMM seeing monitor ranged between 0.7 and 1.0\,arcsec for \cluster\ HSC-CL2330, which was executed first, and between 0.5 and 1.2\,arcsec for \cluster\ HSC-CL2329. Note that the seeing FWHM in the $H$-band used for the KMOS spectroscopy is considerably smaller, with a median around 0.3\,arcsec. 

\begin{figure*}[t]
\centering
\includegraphics[width=15.5cm]{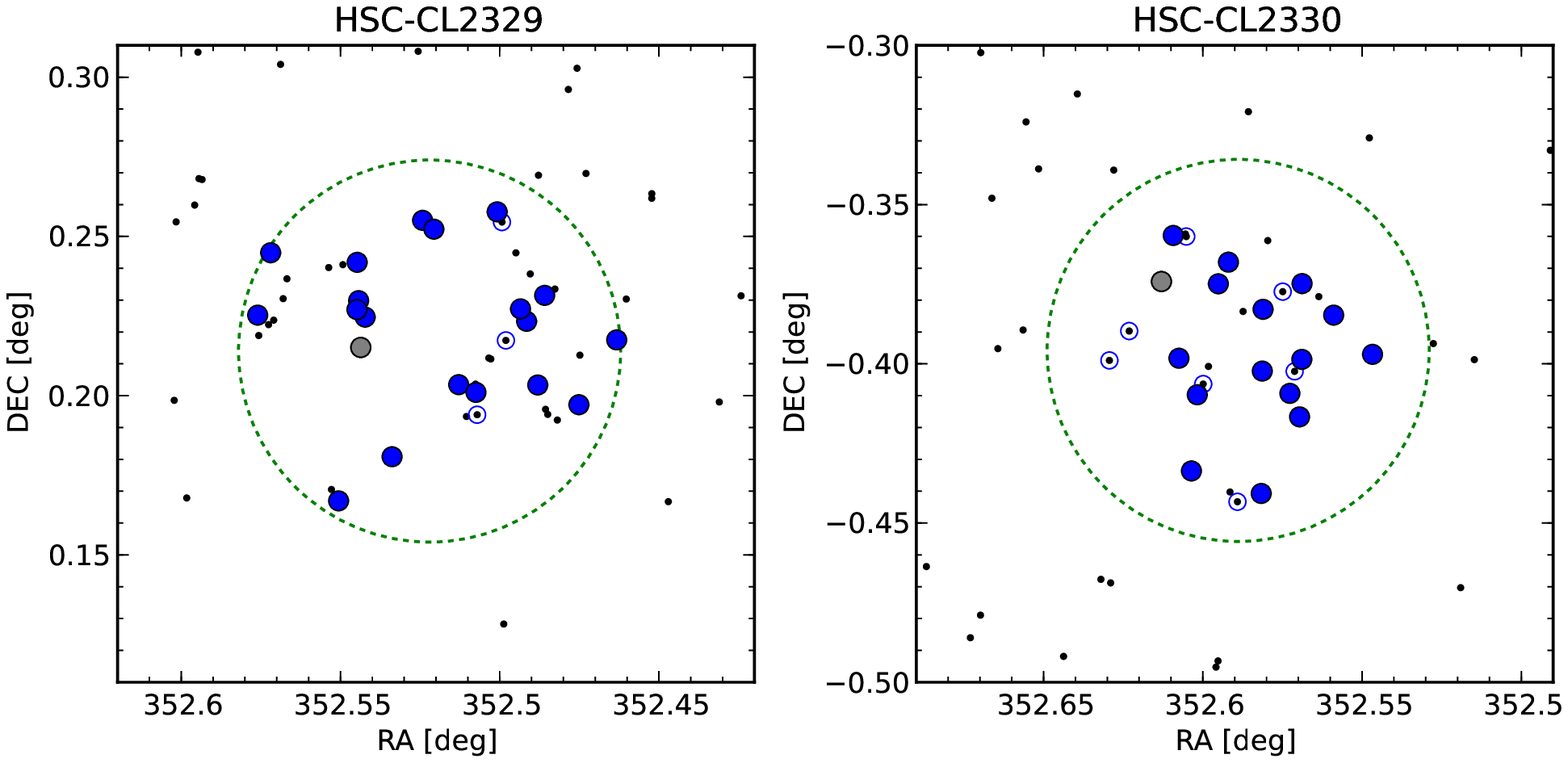}
\includegraphics[angle=270,width=13cm]{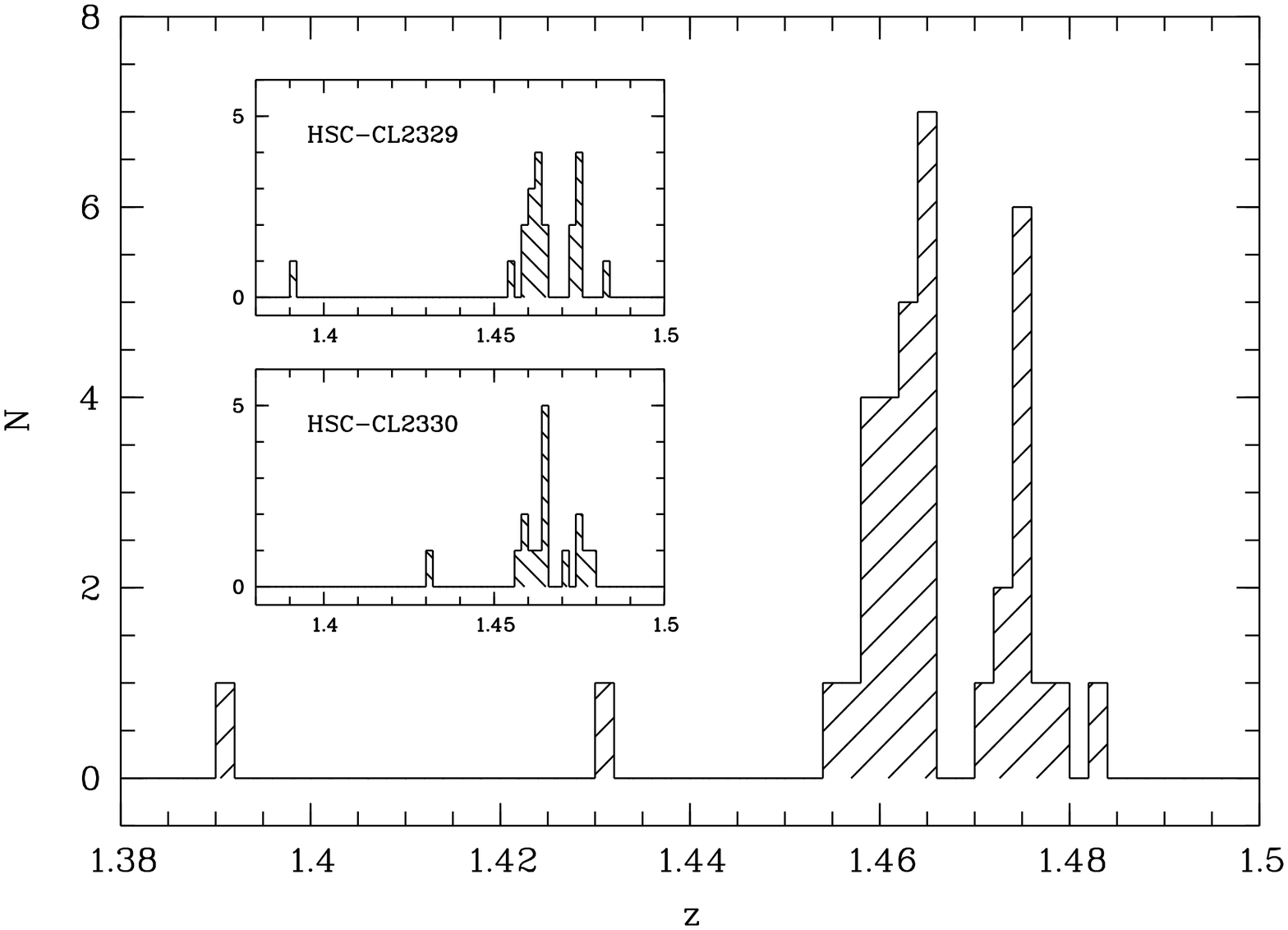}
\caption{\label{zhis} 
\emph{Top:} Spatial distribution of our target candidates (small dots), galaxies without \ha\ detections (open circles), and \ha-detected galaxies (filled circles) for the two observed \clusters. The large dotted circle in both plots denotes the available KMOS patrol field with a diameter of 7.2\,arcmin or $\sim$\,3.7\,Mpc at the \clusters' redshifts. 
\emph{Bottom:}
Redshift distribution of all 36 \ha\ detections for both \clusters\ combined (large plot) and for the two \clusters\ individually (inset graphs). Only two field galaxies are found (shown as grey filled circles in the upper panel), demonstrating the efficiency of the target selection. Both \clusters\ reside at a redshift $z=1.47$, which we aimed for by selecting overdensities of \oii\ emitters at $z \approx 1.5$ with combined broad- and narrow-band data. Interestingly, the two \clusters\ show similar sub-structure in redshift space.
}
\end{figure*}

\section{\label{data}Data Reduction}

The data reduction was carried out with the official ESO-KMOS pipeline version 1.4.3. We conducted a range of tests as to whether the default settings of the pipeline can be improved. In these tests, we aimed to optimize the $S/N$ in the continuum close to the \ha\ line and in the \ha\ line itself, and to minimize night-sky OH residuals mainly $\pm\,200$\,\AA\ of \ha. We decided to deviate from the default reduction settings in two respects.

Firstly, we detected significant offsets between the IFU object positions from the individual Observation Blocks (OBs) of \cluster\ HSC-CL2329, typically only 1-2 pixels but in three cases as large as four pixels (0.8\,arcsec). Although these offsets do not compromise the spatial coverage of our targets, they are much larger than expected based on the IFU-positioning accuracy. They probably are related to a re-calibration of the IFU positioning system during the observation epoch of HSC-CL2329 (M.~Hilker, priv.~comm.). We accounted for these offsets during the co-addition of the individual exposures. 

Secondly, the best data quality in the co-addition of individual exposures is achieved using sigma clipping. The default setting of the pipeline is to first combine exposures from within a single OB before then combining these OBs into final data cubes. The sigma clipping is, however, much more effective when the number of frames used for the final co-addition is maximized, so we modified the default approach and ran the sigma clipping and cube combination on all exposures from all OBs in one go. 

After some further tests, we disregarded the two OBs with the worst seeing of around one arcsec FWHM in the final data cube combination in both \clusters; i.e.~one OB was rejected for each \cluster. Due to this, the average DIMM seeing was 0.7\,arcsec FWHM for HSC-CL2330 and 0.5\,arcsec FWHM for HSC-CL2329, and the total time on target 5040\,s for all galaxies.

\section{\label{ana}Analysis}

We detected \ha\ in 36 out of 46 targets. Only two galaxies are foreground objects probably not physically associated with the \clusters. Our combined narrow-band/photo-$z$ selection strategy thus was confirmed as highly efficient. The redshift distributions and sky positions of our sample are shown in Fig.~\ref{zhis}. Given that the projected separation of the two \clusters\ is approx.~19\,Mpc, their $z$-distributions are remarkably similar with peaks at $z\approx1.46$ and $z\approx1.475$. This probably hints to Large Scale Structure that both \clusters\ are part of.
The gap in the redshift distribution of both \clusters\ (corresponding to $\sim\,16200$\,\AA) is very unlikely to be due to problems with strong night-sky residuals. Three strong OH lines are located in the vicinity of $z\approx1.47$ \ha\ emission: at 16129\,\AA, 16195\,\AA, and 16235\,\AA, with the first and last one being much stronger than the middle one. Among the galaxies with determined redshifts, we have six cases each where the \ha\ line profile was affected by residuals of the line at 16129\,\AA\ or 16235\,\AA, but no spectra where \ha\ is affected by the OH line at 16195\,\AA. The gap in the $z$-distribution of both \clusters\ hence most probably is physical.

\begin{figure}[t]
\resizebox{\hsize}{!}{\includegraphics[angle=270]{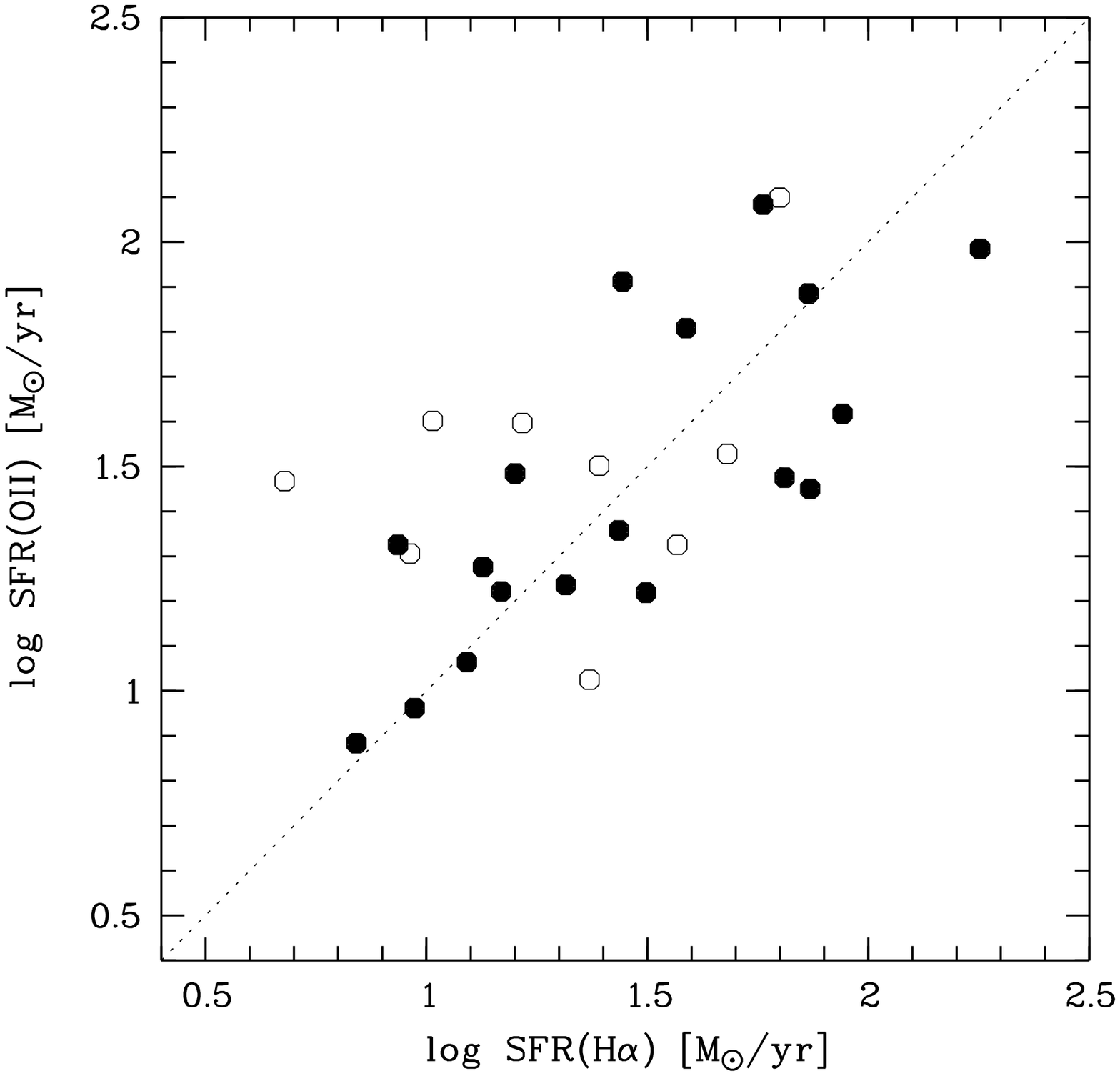}}
\caption{\label{sfrs}
Comparison between the star formation rates from \oii\ fluxes estimated with HSC SSP NB imaging and the new, \ha-based SFRs from the KMOS data. Open symbols depict objects for which the \ha\ line was affected by night sky residuals, which could lead to underestimated SFRs; however, there is no clear indication for this.
}
\end{figure}

12 of the \ha\ detections are weak and extend over only a few spaxels in the data cubes. These data only allow to determine the redshift. In the remaining 24 detections, the \ha\ emission was more spatially extended so that we could use them to extract velocity fields (VFs). 

\ha\ luminosities were transformed into star formation rates (SFRs) following Kennicutt~(\cite{kenn98}). Since the only prominent emission feature in the data is the \ha\ line ($+$\,\nii), the extinction coefficient could not be determined from a traditional source like the Balmer decrement. Instead, we used chemical enrichment models from Ferreras et al.~(\cite{ferr14}) to determine the mass- and disk inclination-dependent $E(B-V)$ extinction. This was then converted into the A(\ha) extinction coefficient following Ly et al.~(\cite{ly12}). In Fig.~\ref{sfrs}, the SFRs computed from \ha\ are compared to the \oii-based ones derived from the NB photometry following Gilbank et al.~(\cite{gilb10}). We find a good agreement between the two (with a scatter of $\sim$\,0.31\,dex), even for cases where the \ha\ line is affected by night sky residuals.

For the determination of the intrinsic maximum rotation velocity $\vm$, synthetic VFs were created based on gas disk inclination, kinematic center, kinematic position angle, turnover radius and maximum rotation velocity for a given galaxy. Intrinsically, the rotation velocity is assumed to rise linearly with radius and turn over into a regime of constant rotation velocity $\vm$ at a turnover radius $r_{\rm t}$ that is linked to the stellar disk scale length $\rd$ via $r_{\rm t} = 2.2\,\rd$. 
This turnover radius, strictly speaking, follows from the surface mass density profile of a pure exponential disk (Binney \& Tremaine~\cite{binn87}) and is often used in the literature to fit observed rotation curves of spiral galaxies (e.g.~Courteau~\cite{cour97}).
The synthetic VFs take into account blurring due to seeing during spectroscopy. The optical DIMM seeing monitor values were transformed into the corresponding $H$-band values following prescriptions provided by ESO. For our data, the $H$-band seeing ranged from 0.25 to 0.3\,arcsec. The VF models also account for beam smearing stemming from the finite pixel size of 0.2\,arcsec. 

VFs covering 30-40 spaxels allow to use all parameters listed above as free fitting parameters. For the VFs with the smallest number of information elements, however, $\vm$ is the only free parameter, while all other parameters were determined from the $i$-band imaging with the GALFIT package (Peng et al.~\cite{peng02}). As a cross-check, we used the output of Source Extractor (Bertin \& Arnouts~\cite{bert96}). 
Position angles and axial ratios $b/a$ were in good agreement between GALFIT and SExtractor for most objects, but in a few cases, very low ratios $b/a<0.1$ were yielded by GALFIT, coinciding with significant fit residuals. This was solved by using a synthetic Gaussian profile as input Point Spread function instead of a PSF constructed from averaged stellar images, which was used for all other objects. For these measurement of structural parameters, we used the best-seeing HSC-SSP images, which are $i$-band frames with a total integration time of 30\,min and a FWHM of 0.7\,arcsec.

\begin{figure*}[t]
\centering
\includegraphics[width=18.5cm]{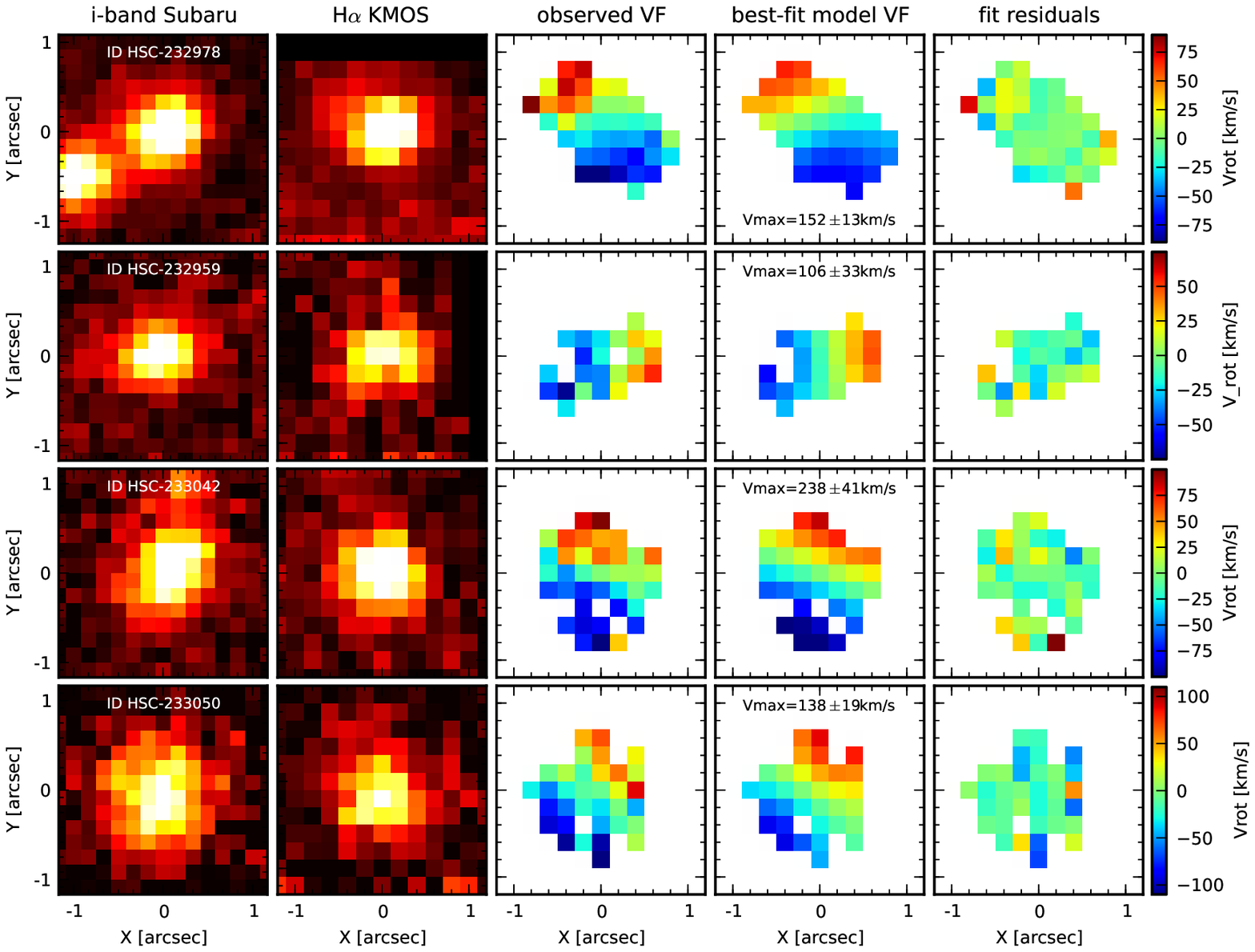}
\caption{\label{vfobs}
Four examples of $z\approx1.47$ \cluster\ galaxies for which the maximum rotation velocity $\vm$ could be determined. 
In each row, the plots show from left to right: best-seeing $i$-band image from the Subaru HSC survey, ionized gas map as observed in H$\alpha$ with VLT/KMOS, observed rotation velocity field, best-fit model rotation velocity field used for the determination of the intrinsic $\vm$, fit residuals after subtracting the best-fit model from the observed velocity field. Note that the galaxy to the southeast of HSC-232978 (top panel, leftmost figure) is a foreground galaxy with a photometric redshift, depending on the method, in the range $0.7<z_{\rm phot}<0.8$, undetected in the KMOS data.
}
\end{figure*}

Our $\vm$ derivation algorithm does not perform a Levenberg-Marquard minimization nor uses the Monte Carlo Markov Chain approach but probes the entire parameter space. This is computationally expensive but avoids running into any local $\chi^2$ minima. 14 velocity fields yielded a robust value for $\vm$, while the remaining 10 VFs could not be properly fitted due to insufficient extent, strong perturbations or a total lack of a velocity gradient. All 14 galaxies stem from the \clusters; neither of the two foreground field galaxies yielded a $\vm$. Four examples of observed and model VFs, along with $i$-band images and \ha\ maps, are shown in Fig.~\ref{vfobs}.

\begin{figure*}[t]
\centering
\includegraphics[width=9cm,angle=270]{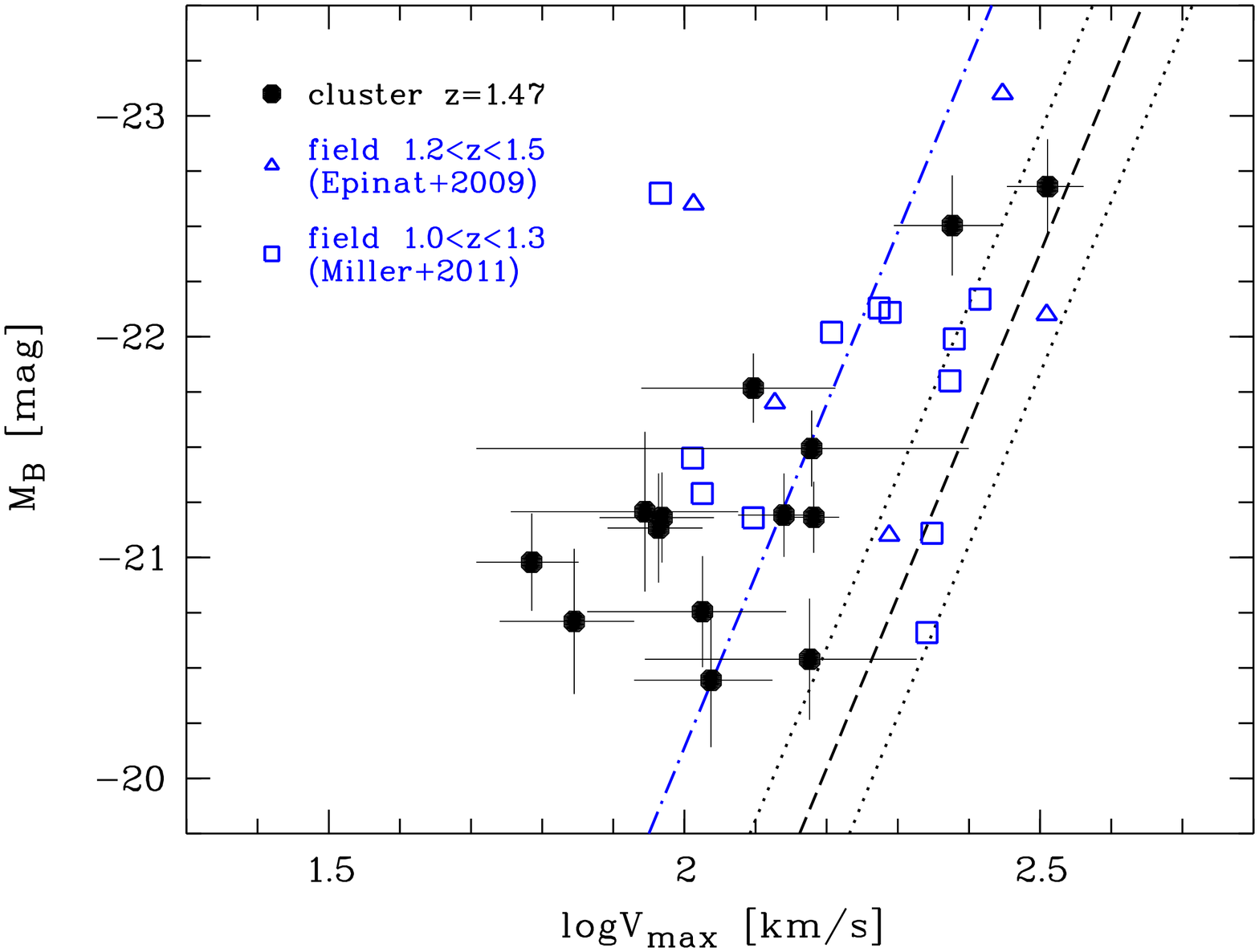}
\caption{\label{tfrb}
$B$-band Tully-Fisher diagram showing our \cluster\ galaxy sample at $z=1.47$ (filled black circles).
We also show field galaxies at $1.0<z<1.3$ (open squares) from Miller et al.~(\cite{mill11}) and  $1.2<z<1.5$ (open triangles) from Epinat et al.~(\cite{epin09}) for comparison.
The local Tully-Fisher relation from Tully et al.~(\cite{tull98}) is depicted by a dashed line; the dotted lines indicate the 1$\sigma$ scatter. 
The dash-dotted line denotes the expected TFR for field galaxies at $z=1.5$, adopting the luminosity evolution found in B\"ohm \& Ziegler~(\cite{boeh16}).
}
\end{figure*}

Rest-frame absolute $B$-band magnitudes were computed from the apparent magnitudes in the $y$-filter which, among the available HSC-SSP filters, best probes the rest-frame $B$-band at the redshifts of the KMOS targets. The $k$-corrections for the transformation $y \rightarrow B$ were derived via synthetic photometry. Intrinsic absorption was taken into account using the inclination- and $\vm$-dependent prescription from Tully et al.~(\cite{tull98}). We also use this work as a local comparison sample to ensure consistency. 

We note that, by adopting the Tully et al. approach up to $z=1.5$, we assume that there is no redshift-dependence of the dust mass fraction at given mass or of the dust properties such as typical grain sizes. Based on the data at hand, we can not infer the amount of intrinsic dust absorption by means of SED fitting (e.g. Wolf et al.~\cite{wolf09}) or Balmer decrement (e.g. Kennicutt et al.~\cite{kenn09}). High-redshift galaxies have higher SFRs, and, in turn, a higher production of dust (e.g. Calzetti~\cite{calz01}). It would therefore be possible that by using the Tully et al.~(\cite{tull98}) method at $z \approx 1.5$, we underestimate the $B$-band luminosities of our KMOS targets and, in turn, the evolution in luminosity compared to $z \approx 0$. On the other hand, e.g. Scoville et al.~(\cite{scov15}) find that the attenuation curve of $2<z<4$ galaxies is similar to that of local starburst galaxies from Calzetti et al.~(\cite{calz00}). Furthermore, Whitaker et al.~(\cite{whit17}) find that the fraction of obscured star formation at given stellar mass shows little evolution over the redshift range $0<z<2.5$.

The $B$-band Tully-Fisher diagram, comprising the $z=1.47$ \cluster\ galaxies, $1<z<1.5$ field spirals (taken from Epinat et al.~\cite{epin09} and Miller et al.~\cite{mill11}), and the local TFR is shown in Fig.~\ref{tfrb}. 
The distant \cluster\ galaxies as well as almost all distant field galaxies fall on the high-luminosity side of the local TFR. Computing the offsets $\Delta M_B$ (for a given $\vm$) from the local relation, we find median values of $\langle \Delta M_B \rangle = -1.87$\,mag for the \cluster\ galaxies and $\langle \Delta M_B \rangle = -1.37$\,mag for the field sample. Given the lower average redshift of the field galaxies ($\langle z \rangle = 1.10$) compared to the \cluster\ sample ($\langle z \rangle = 1.47$), the difference in the TF offsets could be due to the difference in look-back time and the corresponding difference in luminosity evolution. Adopting the redshift-dependent luminosity evolution reported for redshifts $0.1<z<1.0$ in B\"ohm \& Ziegler~(\cite{boeh16}) and projecting it to higher redshifts, we find that the expected TF offsets of the field galaxies at $\langle z \rangle = 1.10$ would be $\Delta M_B = -1.38$\,mag. This is a stronger evolution than predicted from combined $N$-body simulations and semi-analytic models of Dutton et al.~(\cite{dutt11}; $\langle \Delta M_B \rangle = -0.90$\,mag), but in excellent agreement with the observed value $\langle \Delta M_B \rangle = -1.37$\,mag we find for the combined samples of Miller et al. and Epinat et al. For $z=1.5$, the expected luminosity evolution would be $\Delta M_B = -1.65$\,mag. The average TF offset of the \cluster\ sample hence is only slightly larger (by ~0.2\,mag) than what would be expected for field galaxies at $z=1.5$. This indicates that, at least in a statistical sense, the impact of the dense environment on the average luminosity evolution of disk galaxies is small at that epoch.

Note that the stellar-mass TFR is not part of the scope of this paper. The currently available photometry of the HSC-SSP is comprising filters that probe only up to rest-frame wavelengths $\lambda\,\approx\,4300$\,\AA\ for galaxies at $z=1.47$. Stellar mass determinations based on the UV / blue part of the spectral energy distribution carry very large systematic errors. Without NIR imaging at our disposal, we prefer to restrict our TFR analysis to the $B$-band.

As an additional tool in our kinematic analysis, we use a measure for the kinematical asymmetry $A_{\rm kin}$. To this end, we rely on the formalism used in B\"osch et al.~(\cite{boes13a}), which is a variation of the formalism presented by Dale et al.~(\cite{dale01}). It quantifies the asymmetry of the rotation curve (rotation velocity as a function of radius) of a galaxy. 

For a given rotation curve, $A_{\rm kin}$ is computed via:
\begin{equation}  
A_{\rm kin} = \sum_{i}\frac{|v(r_{i})+v(-r_{i})|}{\sqrt{\sigma_{v}^{2}(r_{i})+\sigma_{v}^{2}(-r_{i})}}\cdot \left[ \frac{1}{2}\sum_{i}\frac{|v(r_{i})|+|v(-r_{i})|}{\sqrt{\sigma_{v}^{2}(r_{i})+\sigma_{v}^{2}(-r_{i})}}\right]^{-1}
\label{eq:asym}
\end{equation}
Here, $v(r_{i})$ and $v(-r_{i})$ are the observed rotation velocities at a galactocentric radius $r_i$ on either side from the galaxy's center, while $\sigma_v(r_i)$ and $\sigma_v(-r_i)$ are the errors on the rotation velocity at those positions. The total area between the kinematically folded, approaching and receding side of the rotation curve is normalized to the average area under the rotation curve. Additionally, the contribution of each velocity pair ($v(r_{i}),v(-r_{i})$) is weighted by its error ($\sigma_{v}(r_{i}),\sigma_{v}(-r_{i})$).
The asymmetry $A_{\rm kin}$ is then minimized by varying the kinematic center within $\pm\,2$\,pixels of the photometric center (corresponding to $\sim\,3.4$\,kpc at $z=1.47$).
Numerically, $A_{\rm kin}$ can vary between a value of zero for a perfectly symmetric rotation curve and a maximum value of 2 if all line-of-sight velocities $v(r_i)$ have the same sign at all covered radii $r_i$.

Our main motivation to use rotation curves instead of the full 2-D velocity fields to quantify kinematic asymmetries is to compare the results obtained in the HSC \clusters\ to the cluster system A901/902 at $z=0.17$, for which our team carried out multi-object spectroscopy with VLT/VIMOS (B\"osch et al.~\cite{boes13a}). We extracted rotation curves from the velocity fields of our KMOS data by placing mock slits along the photometric major axis, using an apparent slit width of 0.4\,arcsec. At $z=1.5$, this corresponds to a physical slit width of 3.4\,kpc, which is the same physical width of the slits used in our VIMOS spectroscopy of $z=0.17$ A901/902 galaxies, where the apparent slit width was 1.2\,arcsec. The spatial resolution of the VIMOS observations (taken under an optical FWHM $\sim$1\,arcsec) is, in physical units, very similar to resolution of the KMOS observations ($H$-band FWHM $\sim$0.3\,arcsec), given the difference in redshift. The comparison between the $z=0.17$ and $z=1.5$ data will be carried out in Sect.~\ref{disc}.

We also computed morphological asymmetries $A_{\rm morph}$ on the best-seeing $i$-band images, following the definition of Conselice~(\cite{cons03}). The asymmetry compares the original image to a version of itself that is rotated by 180$^\circ$. To account for contributions by sky noise, a blank sky region $B$ that has the same size as the object image is included in the computation:

\begin{equation}
A_{\rm morph} =  \min \Bigg( \frac{\sum_{i,j} | I - I_{180}|}{\sum_{i,j} |I|} - \frac{\sum_{k,l} | B - B_{180}|}{\sum_{i,j}|I|} \Bigg)
\end{equation}

Here, $I$ is the original image, $I_{180}$ is the image rotated by 180$^\circ$ about the adopted galaxy center; $B$ and $B_{180}$ are the background and rotated background. The sum is computed over all pixels within the 1$\sigma$ isophotes of the galaxy, as determined using Source Extractor. $B$ covers the same number of pixels as $I$.  $A_{\rm morph}$ is minimized by allowing small shifts of the assumed position of the rotation axis (i.e.~the galaxy center), by a maximum of $\pm$\,2 pixels in $x$- or $y$-direction.

We furthermore derived the gas velocity dispersion $\sigma_{\rm g}$ in the disk, by measuring the velocity dispersion in each spaxel (corresponding to $\sim1.7$\,kpc on a side) across the entire velocity field of a given galaxy and then computing its median.
Taking into account the spectral resolution of KMOS in the $H$-band ($R\approx4000$), we found gas disk velocity dispersions in the range 16.8\,km/s\,$\,<\,\sigma_{\rm g}\,<\,$\,33.7\,km/s. Note that, for our data set, the conclusions drawn in the following discussion would not change if the gas velocity dispersion $\sigma_{\rm g}$ instead would be determined in a given galaxy's center.

\begin{table*}[t]
\centering
\caption[]{\label{datatab}The main parameters of the 14 galaxies from our sample for which a derivation of the maximum rotation velocity $\vm$ was feasible.
}
\small
\begin{tabular}{@{}cccccccccc@{}}\toprule
ID & $z$ & $y$ & $M_B$ & $\sigma_{\rm MB}$ & $\vm$ & $\sigma_{\rm vmax}$ & $\sigma_{\rm g}$ & SFR(\ha) &  $A_{\rm kin}$ \\
 & & [mag] & [mag] & [mag] & [km/s] & [km/s] & [km/s] & [$M_\odot/$yr] & \\
\midrule
HSC-232916 & 1.4650 & 23.35 & -20.54 & 0.27 & 150 &  62 & 30.5 &   6.9 & 
0.26 \\	
HSC-232950 & 1.4822 & 22.99 & -21.21 & 0.36 &  88 &  31 & 33.7 &  31.4 & 
0.38 \\	
HSC-232953 & 1.4607 & 22.72 & -21.18 & 0.20 &  93 &  17 & 19.7 &  14.8 & 
0.69 \\	
HSC-232955 & 1.4619 & 23.27 & -20.71 & 0.33 &  70 &  15 & 20.2 &  23.4 & 
0.18 \\	
HSC-232956 & 1.4625 & 22.87 & -20.98 & 0.22 &  61 &  10 & 17.0 &  12.3 & 
0.80 \\	
HSC-232959 & 1.4634 & 23.16 & -20.75 & 0.25 & 106 &  33 & 24.4 &   9.4 & 
0.17 \\	
HSC-232966 & 1.4751 & 22.43 & -21.49 & 0.17 & 151 & 100 & 28.9 &  16.5 & 
0.46 \\	
HSC-232978 & 1.4641 & 22.73 & -21.18 & 0.16 & 152 &  13 & 22.4 &  73.2 & 
0.36 \\	
HSC-233036 & 1.4760 & 21.98 & -22.68 & 0.21 & 324 &  40 & 24.4 & 179.0 & 
0.60 \\	
HSC-233042 & 1.4653 & 22.13 & -22.50 & 0.23 & 238 &  41 & 27.7 &  38.7 & 
0.43 \\	
HSC-233050 & 1.4650 & 22.74 & -21.19 & 0.19 & 138 &  19 & 24.6 &  15.9 & 
0.72 \\	
HSC-233051 & 1.4623 & 22.88 & -21.13 & 0.25 &  92 &  14 & 16.8 &  13.4 & 
0.31 \\	
HSC-233055 & 1.4653 & 22.18 & -21.77 & 0.16 & 125 &  38 & 33.0 &  57.8 & 
1.18 \\	
HSC-233056 & 1.4707 & 23.51 & -20.44 & 0.30 & 109 &  24 & 22.5 &  20.7 & 
1.04 \\ 
\bottomrule
\end{tabular}
\tablefoot{Magnitudes are given in Vega system. $\sigma_{\rm MB}$ and $\sigma_{\rm vmax}$ give the respective errors on the rest-frame $B$-band absolute magnitude $M_B$ and the maximum rotation velocity $\vm$.}
\end{table*}

The main parameters derived in our analysis are given in Table~\ref{datatab}. 

\section{\label{disc}Discussion}

The partly very large TFR over-luminosities we found for the $z = 1.47$ \cluster\ galaxies could be due to several reasons. The high-$z$ galaxies form stars at much higher rates than local galaxies of similar mass, and the higher fraction of young, high-mass stars arising from a higher SFR translates into a lower rest-frame $B$-band mass-to-light ratio $M/L_B$. This effect is the most probable explanation for the redshift-dependent $B$-band TFR offsets at $0<z<1$ discussed in B\"ohm \& Ziegler~(\cite{boeh16}). 

In a dense environment, gravitational (e.g. tidal forces) or hydrodynamical interactions (ram-pressure) can act in addition to the evolution in mass-to-light ratio. Tidal interactions can invoke an increase of the star formation rate, provided that relative velocities between the galaxies are not too high and distances between them not too large (e.g. Lambas et al.~\cite{lamb03}). Moreover, tidal forces can perturb the velocity field and induce non-circular motions of the gas and stars in a galaxy. Both effects (higher SFR and/or a stronger contribution from non-ordered motions) would lead to shifts towards higher luminosity in the Tully-Fisher diagram, as long as the perturbations of the rotation velocity field are not too strong and $\vm$ can still be derived. The effect of ram-pressure, on the other hand, can either increase \emph{or} decrease the luminosity at given $\vm$, depending on whether the interaction is observed at an early or a later stage. 
Simulations have shown that ram-pressure can, via compression of the gaseous disk, initially lead to an enhanced SFR for several 100\,Myr, followed by the quenching of star formation via massive gas loss (e.g.~Ruggiero \& Lima Neto~\cite{rugg17}).

Based on FORS slit spectroscopy, our group previously carried out a Tully-Fisher analysis of cluster disk galaxies at slightly lower redshift $z\,\approx\,1.4$ (P\'erez-Mart\'inez et al.~\cite{pere17}). There, we targeted the massive cluster XMMU J2235-2557, which has a well-established intra-cluster medium as confirmed with X-ray observations (Mullis et al.~\cite{mull05}). 
The average offset from the local $B$-band TFR found by P\'erez-Mart\'inez et al. ($\sim-1.6$\,mag) is similar to what we report here ($\sim-1.8$\,mag), and also the scatter of the TF offsets is similar between the two samples (1.0\,mag at $z=1.4$ vs. 1.1\,mag at $z=1.5$). Moreover, the average luminosity evolution observed in the cluster/\cluster\ environment is only slightly stronger (by $\sim$\,0.2\,mag) than what is expected in the field at the same redshifts (adopting Eq.~4 from B\"ohm \& Ziegler~\cite{boeh16}).
This is in agreement with earlier studies such as Ziegler et al.~(\cite{zieg03}) which have found that, provided that only undisturbed velocity fields / rotation curves are utilized in a Tully-Fisher analysis, little difference is found between dense environments and the field on average.

However, some individual HSC \cluster\ galaxies have much stronger TF offsets~--- up to four magnitudes in rest-frame $B$~--- than what was found in J2235-2557 cluster, where the largest offsets are $\Delta M_B \approx -2.5$\,mag. This suggests that there are other factors having an impact on the Tully-Fisher analysis of the \cluster\ galaxies, e.g. non-circular gas motions; we will address this again further below.

\begin{figure}[t]
\resizebox{\hsize}{!}{\includegraphics[angle=270]{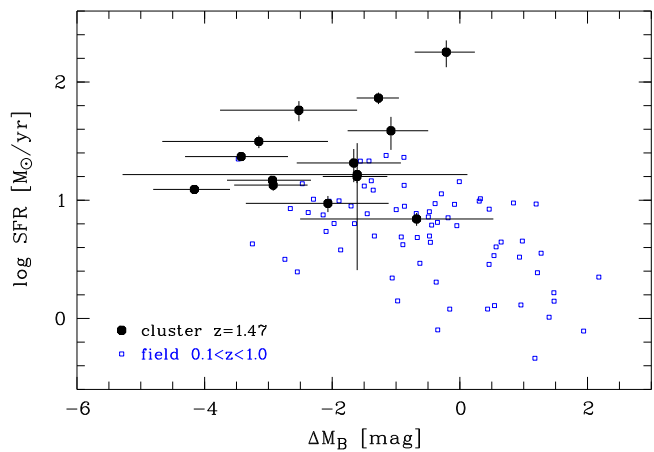}}
\caption{\label{dMBsfr}
\ha-based star formation rates versus offsets $\Delta\,M_B$ from the local $B$-band Tully-Fisher relation. In contrast to field disk galaxies at $z<1$ (small open squares; taken from B\"ohm \& Ziegler~\cite{boeh16}), the TFR offsets of the $z=1.47$ \cluster\ galaxies (filled circles) are not correlated with star formation rate. See text for details.
}
\end{figure}

To investigate the influence of star-formation on the distribution in Tully-Fisher space for our KMOS sample, we show the TFR offsets $\Delta\,M_B$ plotted against SFR in Fig.~\ref{dMBsfr}, compared to the field sample from B\"ohm \& Ziegler~(\cite{boeh16}). For field spirals at $z<1$, there is a clear correlation between the TFR offsets and SFRs, a Spearman test yields $\rho=-0.46$ and $p=10^{-4}$.
For the \cluster\ galaxies, however, we find no clear correlation: $\rho=0.27$ and $p=0.35$. The galaxy with the highest SFR of over 100\,M$_\odot$\,yr$^{-1}$ is a slight outlier in the KMOS distribution, but even if one neglects this object in the statistics, the Spearman test result does not change significantly: for the remaining 13 galaxies, we find $\rho=0.09$ and $p=0.78$. We note that the SFRs also are not correlated with the kinematical asymmetry $A_{\rm kin}$ ($\rho=0.19$ and $p=0.37$), it is hence unlikely that star formation in many of the galaxies is enhanced by tidal interactions.

\begin{figure}[t]
\resizebox{\hsize}{!}{\includegraphics[angle=270]{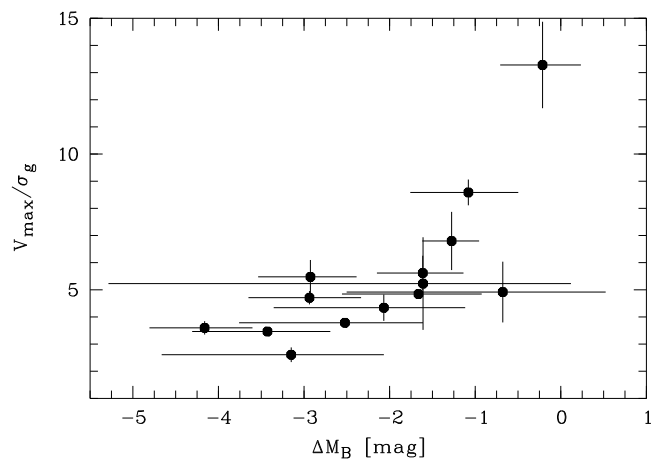}}
\caption{\label{dMBvsig}
Offsets from the local Tully-Fisher relation $\Delta\,M_B$ versus ratio between maximum rotation velocity $\vm$ and gas velocity dispersion $\sigma_{\rm g}$. Galaxies which are more rotation-dominated show smaller TFR offsets. This indicates that significant non-circular motions in the gas kinematics are contributing to the deviations from the local TFR.
}
\end{figure}

The KMOS sample has some overlap in SFR and $\Delta\,M_B$ with the highest-SFR field galaxies at $z<1$, but the above tests indicate that star formation rate is not the only parameter that drives the offsets of the $z \approx 1.5$ galaxies from the local TFR. 
Indeed, we find that the TFR offsets depend on the contributions of non-circular motions to the $z=1.47$ \cluster\ galaxies' gas kinematics. Fig.~\ref{dMBvsig} shows the TFR offsets versus the ratio $\vm/\sg$ between maximum rotation velocity and gas velocity dispersion. This figure demonstrates that galaxies with lower $\vm/\sg$ ratios (less rotation-dominated kinematics) have larger TFR offsets, confirmed by a Spearman test which returns $\rho=0.77$ and $p=0.001$.
A relatively large velocity dispersion does not necessarily imply gravitational perturbations e.g.~by galaxy-galaxy interactions, but could also indicate kinematically ``hotter'' disks that have also been found in the field regime at similar redshifts (e.g.~Simons et al.~\cite{simo16}). We can, however, not compare this result to our own $z<1$ field sample as we did in Fig.~\ref{dMBsfr}, because the spectral resolution of those data, which were taken with VLT/FORS, is too low; the spectral resolution of the FORS grism used in that campaign corresponds to a lower limit $\sg \approx 100$\,km/s.

\begin{figure}[t]
\resizebox{\hsize}{!}{\includegraphics[angle=270]{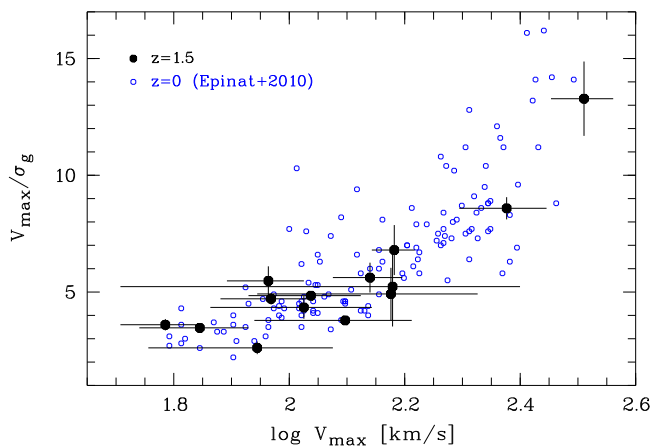}}
\caption{\label{vmvsig}
Ratio between maximum rotation velocity $\vm$ and gas velocity dispersion $\sigma_{\rm g}$ as a function of maximum rotation velocity for our \cluster\ sample at $z=1.5$ (filled circles) and local disk galaxies from Epinat et al.~(\cite{epin10}, open circles). In both samples, faster rotators (i.e.~disk galaxies of higher mass) have more rotation-dominated kinematics that galaxies of lower mass. 
}
\end{figure}

Independent of whether the low $\vm/\sg$ ratios of some objects are due to gravitational interactions or a general trend towards kinematically hotter, thicker disks at higher redshifts (e.g.~Wisnioski et al.~\cite{wisn15}), we would expect more massive galaxies to be more rotation-dominated than lower-mass ones: higher-mass galaxies are less susceptible to external gravitational forces, and they also settle to high $\vm/\sg$ ratios at earlier cosmic times than low-mass disks (e.g.~Kassin et al.~\cite{kass12}).
Fig.~\ref{vmvsig} shows the ratio $\vm/\sg$ between maximum rotation velocity and gas velocity dispersion as a function of $\vm$, which is a proxy for total mass, for our $z=1.5$ \cluster\ sample and, in comparison, local field galaxies from the GHASP survey (Epinat et al.~\cite{epin10}). This survey comprises Fabry-P\'erot \ha\ observations of local isolated spiral galaxies. For the purpose of this plot, we have restricted the local sample to the $\vm$ range covered by the HSC \cluster\ galaxies. In both samples, the kinematics of galaxies with higher $\vm$, hence higher mass, are more rotation-dominated than those of low-mass ones.
However, the $\vm/\sg$ ratio is slightly lower for the distant \cluster\ galaxies; we find a median $\langle \vm/\sg \rangle = 4.9$ compared to $\langle \vm/\sg \rangle = 6.2$ in the GHASP sample. This could reflect the increase of gas velocity dispersion with look-back time (e.g.~Wisnioski et al.~\cite{wisn15}) or velocity field perturbations due to galaxy-galaxy interactions in the $z=1.5$ \cluster\ environment. As we discuss further below, the relative velocities of the \cluster\ galaxies in our sample are probably much lower than would be typical in low-redshift clusters.
From a theoretical perspective, besides more frequent gravitational interactions, several scenarios exist to explain the increasing gas velocity dispersion in disks towards higher redshifts. Hung et al.~(\cite{hung19}) used cosmological simulations to show that higher gas inflow rates and subsequently enhanced star formation rates lead to larger $\sg$ at earlier cosmic times.

\begin{figure}[t]
\resizebox{\hsize}{!}{\includegraphics[angle=270]{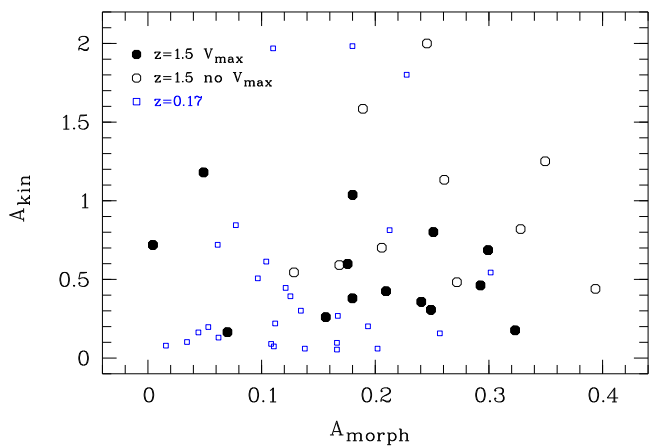}}
\caption{\label{am_avf}
Morphological asymmetry $A_{\rm morph}$ versus kinematic asymmetry $A_{\rm kin}$ for our $z=1.5$ \cluster\ sample.
Filled circles denote galaxies with determined $\vm$, open circles represent galaxies with disturbed or non-rotating velocity fields that did not allow to derive $\vm$. For comparison, we also show spirals from the low-redshift cluster A901/902 at $z=0.17$ (small open squares), using data from B\"osch et al.~(\cite{boes13a}). The imaging of the $z=0.17$ galaxies has been degraded to match the image quality of the $z=1.5$ data, and also the derivation of $A_{\rm kin}$ ensured comparable spatial resolutions in the two data sets.
}
\end{figure}

We now want to include morphological information in our analysis, in an attempt to disentangle various possible interaction processes in the high-redshift \clusters. In Fig.~\ref{am_avf}, we show the kinematical asymmetry $A_{\rm kin}$ as a function of morphological asymmetry $A_{\rm morph}$. 
The Y-axis of this parameter space is sensitive both to hydrodynamical and gravitational interactions, while the X-axis only probes gravitational interactions; stars are not affected by ram-pressure (e.g., Kronberger et al.~\cite{kron08}).
We first want to focus on the $z=1.5$ \cluster\ data. Filled circles denote galaxies with a determined $\vm$, while open circles show galaxies with disturbed/non-rotating velocity fields, for which $\vm$ could not be derived.
As would be expected, galaxies with velocity fields that could be used for the determination of $\vm$ generally have lower kinematic asymmetries $A_{\rm kin}$ 
(median $\langle A_{\rm kin} \rangle = 0.44$)
than galaxies that did not yield a value for $\vm$ 
($\langle A_{\rm kin} \rangle = 0.76$). 
No correlation between $A_{\rm kin}$ and $A_{\rm morph}$ is found using a Spearman test: $\rho=0.05$ and $p=0.79$.

For the further interpretation, we now will compare to a dense environment at low redshift, the cluster system A901/902 at $z \approx 0.17$.
In this system, we previously found a population of disk galaxies that, despite having regular morphologies with low $A_{\rm morph}$ values, show high gas kinematical asymmetries (cf.~Fig.~17 in B\"osch et al.~\cite{boes13a}). This is evidence for hydrodynamical interaction, as this is the sole process in dense environments that only affects the gas distribution and gas kinematics, but not the stellar light morphology.
To be able to directly compare the A901/902 data to our new $z=1.5$ observations, we have degraded the HST/ACS imaging of galaxies in A901/902 to match the physical pixel scale (1.7\,kpc/pix), spatial resolution (PSF FWHM of 5.9\,kpc) and noise properties of the HSC $i$-band imaging of the $z=1.5$ \clusters. The morphological asymmetries $A_{\rm morph}$ of the $z=0.17$ cluster galaxies were then re-computed on these degraded images. As explained in the last section, we also have extracted rotation curves and derived kinematical asymmetries $A_{\rm kin}$ from the KMOS data cubes such that this parameter is directly comparable between the low- and high-redshift data. The A901/902 sample was restricted to star-forming galaxies with stellar masses $\log (M_\ast/M_\odot) > 9.5$, which is the expected stellar mass range of the KMOS sample, given its distribution of $\vm$ values and the observed stellar mass TFR at $z\approx1.5$ reported by \"Ubler et al.~(\cite{uebl17}). This mass-restricted part of the A901/902 sample is depicted in Fig.~\ref{am_avf} by open squares. 

As is the case for the $z=1.5$ \cluster\ galaxies, no correlation between $A_{\rm morph}$ and $A_{\rm kin}$ is found in the degraded A901/902 sample (Spearman test: $\rho=0.12$ and $p=0.54$). Moreover, the strongly limited spatial resolution of the degraded imaging camouflages the hydrodynamical interactions reported in B\"osch et al.~(\cite{boes13a}): dividing the $z=0.17$ sample into two equally-sized parts with low and high morphological asymmetry ($A_{\rm morph}<0.12$ and $A_{\rm morph}>0.12$, resp.), the low-asymmetry sample does not yield a correlation between $A_{\rm morph}$ and $A_{\rm kin}$ ($\rho=0.19$ and $p=0.51$), and, in particular, no \emph{anti-}correlation as was found based on the fully resolved imaging (cf.~B\"osch et al.~\cite{boes13a}). We hence conclude that hydrodynamical interactions can not be identified at the resolution of the HSC imaging of the \clusters, at least at our current sample size.

Nevertheless, we do find that both the morphological and kinematical asymmetries are higher on average in the $z=1.5$ sample (median $\langle A_{\rm morph} \rangle = 0.23$ and $\langle A_{\rm kin} \rangle = 0.60$) than in the $z=0.17$ data ($\langle A_{\rm morph} \rangle = 0.12$ and $\langle A_{\rm kin} \rangle = 0.22$), probably due to much more efficient galaxy-galaxy interactions in the \clusters\ arising from the low relative velocities between the galaxies, as we will show further below.
We note that part of the difference in average morphological asymmetries between the A901/902 and the \cluster\ galaxies might stem from the difference in rest-frame wavelength of the imaging ($\sim$\,5100\,\AA\ for A901/902, taken with the F606W filter of HST/ACS, versus $\sim$\,3200\,\AA\ for the $z=1.5$ galaxies; $i$-band data).
The sensitivity of the NUV to clumpy star-forming regions generally leads to higher $A_{\rm morph}$ than rest-frame optical or NIR images (that are currently not at hand for the \clusters\ sample).
Irrespective of this, the lack of a correlation between morphological and kinematical asymmetry hints that a mix of hydrodynamical and gravitational interactions might be present in HSC-2329 and HSC-2330 (as in A901/902), but the limited spatial resolution of the imaging at hand does not allow to disentangle these processes.

We estimated the galaxy velocity dispersions in the two 
\clusters, assuming that the two peaks in their redshift distributions are physical, since it seems unlikely that strong night-sky residuals are the cause of the observed ``gap" at $z=1.468$ in both \clusters. For the two peaks in the $z$-distribution of HSC-CL2329, we find velocity dispersions of $\sigma_{\rm v}=304$\,km/s and $\sigma_{\rm v}=364$\,km/s; the distribution in HSC-CL2330 yield values of $\sigma_{\rm v}=385$\,km/s and $\sigma_{\rm v}=361$\,km/s, resp.

If all structures were virialized, these velocity dispersions would be equivalent to virial masses of $1.6 \times 10^{14}\,M_\odot < M_{\rm vir} < 3.2 \times 10^{14}\,M_\odot$. These estimates follow Crook et al.~(\cite{croo07}) and are based on the observed velocity dispersions and projected pairwise galaxy-galaxy separations of the \cluster\ members.
However, it is much more likely that these numbers are overestimating the \cluster\ masses due to their non-virialized state. The large average pairwise galaxy-galaxy separations in the observed overdensities result in projected virial radii between 1.4 and 2\,Mpc for the four structures. This is very large given the observed range in velocity dispersions: following Finn et al.~(\cite{finn05}), the $\sigma_{\rm v}$ values given above, under the assumption of virialization, would correspond to virial radii of only $0.33\,{\rm Mpc}<R_{\rm 200}<0.42\,{\rm Mpc}$.

Rather than being virialized, it is hence much more likely that the targeted overdensities are in the process of formation. To reflect this in our nomenclature, we refer to them as \clusters\ throughout this paper. A further characterization of HSC-CL2329 or HSC-CL2330 is difficult on the basis of our current data. No prominent Brightest Cluster Galaxy is observed in either \cluster, which is in compliance with an early dynamical state, but we can not test for the presence/absence of a red sequence with the current imaging that probes only rest-frame UV/blue. Since we are lacking deep X-ray data, a direct detection of the ICM is also not feasible.

Irrespective of the dynamical status of HSC-CL2329 and HSC-CL2330, the $\sigma_{\rm v}$ values show that the relative velocities between the \cluster\ galaxies in our sample are low compared to those observed in virialized clusters, and tidal interactions between the galaxies hence are much more efficient than would be expected in a typical cluster environment at later cosmic epochs. This strengthens the interpretation that galaxy-galaxy interactions contribute to the distribution shown in Fig.~\ref{am_avf}, though we would expect a clear correlation between morphological and kinematical asymmetry if tidal forces  would be the \emph{only} interaction process at act.

To summarize, our morpho-kinematical analysis of galaxies in dense environments at $z\,\approx\,1.5$, corresponding to a look-back time of $\sim$\,9\,Gyr, has yielded some similarities to the field galaxy population at this epoch, in particular with regard to a high fraction of disks with significant contributions by non-circular motions. However, two points have to be stressed. Firstly, that our sample stems from a cosmic phase that represents the \emph{onset} of environmental influence on galaxy evolution, and not the clear impact observed at lower redshifts. Secondly, the complex redshift distributions of HSC-2329 and HSC-2330 indicate that both structures are in an early, non-virialized stage, and might lack a dense intra-cluster medium at the time of observation, which would weaken the effects from hydrodynamical interactions.
More IFU observations of galaxies in clusters, including more evolved structures at high redshifts, will be necessary to shed more light on the impact of environment on galaxy evolution at early cosmic stages.

\section{\label{concl}Summary}

We used VLT/KMOS to take Integral Field Spectroscopy of the \ha\ emission in 46 galaxies in two \clusters\ at redshift $z=1.47$ detected as overdensities of \oii\ emitters in HyperSuprimeCam Strategic Survey Program data. In the KMOS data cubes, the \ha\ line was detected in 36 galaxies, of which 34 are \cluster\ members; only two objects are located in the field, demonstrating the efficiency of our selection strategy. 24 galaxies show spatially extended \ha\ emission from which velocity fields could be extracted, while in the remaining 12 objects, the \ha\ emission is spread over only a few pixels and/or is affected by strong residuals of OH night sky lines. 

By fitting the observed velocity fields with simulated velocity fields that take into account geometrical effects like disk inclination and position angle as well as seeing and beam smearing, we could derive the maximum rotation velocity $\vm$ of 14 galaxies. The velocity fields of the remaining 10 galaxies with extended \ha\ are disturbed or non-rotating. We computed quantitative morphological and kinematical asymmetries to compare the $z=1.47$ \cluster\ galaxy data to our own studies of the galaxy population in low-redshift clusters. 

Our main findings are:
\begin{enumerate}
\item Based on the Tully-Fisher relation (TFR), all \cluster\ disk galaxies at $z \approx 1.5$ are more luminous than local spirals at given $\vm$, by up to $\sim$\,4\,mag in rest-frame $B$. However, the \emph{average} evolution in $B$-band luminosity between $z=1.5$ and $z=0$ does only weakly depend on environment, since we find little difference in the average offsets from the local TFR between the field and the \clusters.
We did not consider the stellar-mass TFR here as the currently available photometry does not cover rest-frame red optical or NIR colors mandatory for the derivation of robust stellar masses.
\item The deviations from the local TFR are not correlated with star formation rate, but with the ratio between $\vm$ and gas velocity dispersion $\sg$. In turn, this ratio $\vm/\sg$ is larger towards higher $\vm$, i.e., higher total masses. Besides possible gravitational interactions between \cluster\ members, this might reflect that many low-mass disks have not yet settled to purely rotational kinematics by $z \approx 1.5$. This has also been observed in the field at this epoch.
\item Tests with degraded low-redshift cluster galaxy data show that we can not detect purely hydrodynamical interactions within the limitations of the ground-based HSC imaging. Gravitational interactions, on the other hand, are likely to be more efficient in the \clusters\ than in dense environments at low redshifts, due to rather small relative velocities between the $z=1.5$ \cluster\ members. This would explain the higher kinematical asymmetries we find at $z=1.5$ in comparison to cluster galaxies at low redshift.
\end{enumerate}

\begin{acknowledgements}
BLZ is grateful for an NAOJ visitorship in 09/2018. This publication is supported by the Austrian Science Fund (FWF).
\end{acknowledgements}

%
%

\end{document}